\documentclass[twocolumn]{aastex631}

\newcommand\vsini{v\sin{i}}
\newcommand\kms{\rm km\, s^{-1}}
\usepackage[]{comment}

\usepackage{soul}
\usepackage{xcolor}

\def   \paul{\color{black}}

\begin{document}

\title{Rotation and H$\alpha$ emission in a young SMC cluster: a spectroscopic view of NGC~330}

\correspondingauthor{Paul I. Cristofari}
\email{paul.cristofari@cfa.harvard.edu}

\author[0000-0003-4019-0630]{Paul I. Cristofari}
\affiliation{Center for Astrophysics $\vert$ Harvard \& Smithsonian,
60 Garden Street,
Cambridge, MA 02138, USA}

\author[0000-0002-8985-8489]{Andrea K. Dupree}
\affiliation{Center for Astrophysics $\vert$ Harvard \& Smithsonian,
60 Garden Street,
Cambridge, MA 02138, USA}

\author[0000-0001-7506-930X]{Antonino P. Milone}
\affiliation{Dipartimento di Fisica e Astronomia ``Galileo Galilei'', Univ. di Padova, Vicolo dell’Osservatorio 3, Padova, IT-35122}
\affiliation{Istituto Nazionale di Astrofisica - Osservatorio Astronomico di Padova, Vicolo dell’Osservatorio 5, Padova, IT-3512}

\author[0000-0003-2496-1925]{Matthew G. Walker}
\affiliation{McWilliams Center for Cosmology, Carnegie Mellon University, 5000 Forbes Ave, Pittsburgh, PA 15213, USA}

\author[0000-0002-3856-232X]{Mario Mateo}
\affiliation{Department of Astronomy, University of Michigan, Ann Arbor, MI 48109, USA}

\author[0000-0002-4442-5700]{Aaron Dotter}
\affiliation{Department of Physics and Astronomy, Dartmouth College, 6127 Wilder Laboratory, Hanover, NH 03755, USA}

\author[0000-0002-4272-263X]{John I. Bailey III}
\affiliation{California Institude of Technology - Caltech Optical Observatories, Pasadena, CA 91125, USA}

% \collaboration{20}{(AAS Journals Data Editors)}

% \author{F.X Timmes}
% \affiliation{Arizona State University}
% \affiliation{AAS Journals Associate Editor-in-Chief}

% \author{Amy Hendrickson}
% \altaffiliation{AASTeX v6+ programmer}
% \affiliation{TeXnology Inc.}

% \author{Julie Steffen}
% \affiliation{AAS Director of Publishing}
% \affiliation{American Astronomical Society \\
% 1667 K Street NW, Suite 800 \\
% Washington, DC 20006, USA}



\begin{abstract}
We present an analysis of high-resolution optical spectra recorded for 30 stars of the split extended main-sequence turnoff (eMSTO) of the young ($\sim$\,40\,Myr) Small Magellanic Cloud (SMC) globular cluster NGC\,330. Spectra were obtained with the 
M2FS and MIKE 
spectrographs  located on the Magellan-Clay 6.5\,m telescope. These spectra revealed the presence of Be stars, 
occupying primarily the cool side of the split main sequence (MS). Rotational velocity ($\vsini$) measurements for most of the targets are consistent with the presence of two populations of stars in the cluster: one made up of rapidly rotating Be stars ($<\!\vsini\!> \approx 200$\,$\kms$), and {the other} consisting of warmer stars with slower rotation ($<\!\vsini\!>\approx50$\,$\kms$). 
Core emission in the H$\delta$  photospheric lines was observed for most of the H$\alpha$ emitters. The shell parameter computed for the targets in our sample indicate that most of the observed stars should have inclinations below 75$^{\circ}$.
These results confirm the detection of Be stars obtained through photometry, but also reveal the presence of narrow H$\alpha$ and H$\delta$ features for some targets that cannot be detected with low-resolution spectroscopy or photometry. Asymmetry variability of H$\alpha$ line profiles on the timescales of a few years is also observed, and could provide information on the geometry of the decretion disks.
Observations revealed the presence of nebular H$\alpha$ emission, strong enough in faint targets to compromise the extraction of spectra and to impact narrow band photometry used to assess the presence of H$\alpha$ emission.

\end{abstract}



\section{Introduction} \label{sec:intro}

Over the last decades, high precision photometric measurements have provided evidence of an extended main sequence turnoff (eMSTO) and split main sequence (MS) in the color-magnitude diagrams (CMD) of intermediate-age~\citep{milone-2009} and young globular clusters~\citep[e.g.,][]{milone-2013, milone-2017, correnti-2017}.
The discovery of the eMSTO and split MS indicates the presence of multiple populations in globular clusters with possible variations in ages, metallicities, or rotation.
Different scenarios have been offered to explain those observations, some proposing that multiple generations of stars formed in those clusters, or postulating an extended star formation period on scales of a few 100\,Myrs~\citep{mackey-2008, milone-2009, conroy-2011, keller-2011, goudfrooij-2014}, others proposing that stellar rotation plays a major role in the appearance of multiple stellar populations~\citep{bastian-2009, dantona-2017}. None appear to account for all aspects of the observations~\citep{bastian-2018}. The detailed characterization of stellar populations  within globular clusters remains, therefore, essential to unveil the origin of the observed eMSTO and split MS.

 Recent photometric and spectroscopic studies have suggested that the split MS likely results from two distinct populations of stars: one made up of B stars rotating close to the breakup velocity and surrounded by a decretion  disk responsible for emission lines (Be stars), the other composed of hotter B stars with slower rotation~\citep{dantona-2015, milone-2016, bastian-2017, dupree-2017}. The presence of fast-rotating Be stars was confirmed by the detection of H$\alpha$ emission, believed to be directly linked to the presence of Be stars~\citep{rivinius-2013}, via high-resolution spectroscopic measurements in NGC~1866~\citep{dupree-2017}. Those results further suggested that a difference in rotation velocities should be observed between the two populations within a same cluster, although the rotation velocities of Be stars could not be estimated from H$\alpha$ emission lines.

The current work focuses on the young cluster NGC\,330 of the Small Magellanic Cloud (SMC) whose age was reported to be between 30 and 45 Myr~\citep{keller-2000, milone-2018, eldridge-2020, patrick-2020}, and metallicity is expected to be $\rm [Fe/H]\lesssim-1.0$~dex~\citep[e.g.,][]{piatti-2019}. 
Previous photometric studies of the cluster have reported a large number of Be stars~\citep{milone-2018}, consistent with the works of ~\citet{wisniewski-2006}.~\citet{iqbal-2013}  proposed that a low-metallicity environment may produce a relatively large fraction of Be stars. A study of the multiplicity of NGC\,330 from low-resolution spectra~\citep[R$<$4000,][]{ bodensteiner-2021} reported binary fractions in different regions of the color magnitude diagram (CMD), {\paul finding in particular a  binary fraction of $6$\,\% within the Be stellar population. Such results provide} additional constraints for the previously mentioned scenarios.
A recent study published rotational velocity estimates for B and Be stars in NGC\,330 obtained from low-resolution (R$<$4000) MUSE spectra~\citep{bodensteiner-2023}. The work presented here focuses on stars observed in the same cluster, but relies on high-resolution (R$>$25000) spectra, permitting us to probe the profiles of the observed spectral lines.

In this paper, we present the results of a study of high resolution spectra recorded for dozens of targets in NGC\,330. We targeted the H$\alpha$ line in order to assess the presence of emission features, and a He I line at 4143.76\,\r{A} in order to estimate the rotational velocity of the targets. The spectroscopic material and reduction of the data are introduced in Sec.~\ref{sec:observations} and the results are presented in Sec~\ref{sec:results}. Those results are then discussed in Sec.~\ref{sec:discussion}.

\begin{table*}
	%	\caption{List of M2FS targets, magnitudes, H$\alpha$ line shape (emission e, or absorption a), and projected rotational velocities ($v\sin{i}$) measured by fitting \texttt{TLUSTY} models to the He I line at 4143.76\,\r{A} and the H$\delta$ line. 
		%		%		Values found to be suspiciously large are marked with a star and not considered when computing the average velocity of the population.
		%	}
	\centering
	\caption{Targets observed with M2FS and their parameters.}
	\label{tab:list-m2fs}
	\begin{tabular}{cccccccccc}
		\hline
		Target & R.A. (2000.0) [hh:mm:ss] & Decl. (2000.0) [dd:mm:ss] & $m_{\rm F814W}$\tablenotemark{a} & $m_{\rm F336W}$\tablenotemark{a} & Color\tablenotemark{b} & H$\alpha$ excess\tablenotemark{c} & H$\alpha$\tablenotemark{d} & $\vsini$\tablenotemark{e} & shell\tablenotemark{f}\\
		\hline
		2 & 14.067688 [00:56:16.25] & -72.461457 [-72:27:41.24] & 16.412 & 15.391 & -1.020 & -0.724 & e & 296 & 1.2  \\ 
		3 & 14.082867 [00:56:19.89] & -72.460028 [-72:27:36.10] & 15.417 & 14.118 & -1.299 & -1.692 & e & 185 & ...  \\ 
		5 & 14.066931 [00:56:16.06] & -72.466429 [-72:27:59.15] & 16.071 & 14.757 & -1.314 & -1.844 & e & 203 & ...  \\ 
		7 & 14.095990 [00:56:23.04] & -72.465020 [-72:27:54.07] & 15.133 & 13.924 & -1.210 & -1.665 & e & 98 & 1.1  \\ 
		18 & 14.051482 [00:56:12.36] & -72.463192 [-72:27:47.49] & 17.335 & 16.332 & -1.002 & -1.695 & e & ... & 1.7  \\ 
		33 & 14.107634 [00:56:25.83] & -72.458606 [-72:27:30.98] & 17.097 & 16.038 & -1.059 & -1.643 & e & 398 & 1.1  \\ 
		58 & 14.059443 [00:56:14.27] & -72.475078 [-72:28:30.28] & 15.726 & 14.689 & -1.037 & -2.085 & e & 292 & ...  \\ 
		61 & 14.043880 [00:56:10.53] & -72.454236 [-72:27:15.25] & 17.138 & 15.658 & -1.480 & -0.327 & e & ... & ...  \\ 
		64 & 14.070223 [00:56:16.85] & -72.449696 [-72:26:58.91] & 17.430 & 16.395 & -1.035 & -0.965 & e & ... & $>$2.0  \\ 
		68 & 14.106100 [00:56:25.46] & -72.451997 [-72:27:07.19] & 15.366 & 14.213 & -1.152 & -1.727 & e & 55 & 1.4  \\ 
		73 & 14.027701 [00:56:06.65] & -72.460096 [-72:27:36.35] & 14.389 & 13.079 & -1.309 & -1.568 & e & 168 & ...  \\ 
		75 & 14.129855 [00:56:31.17] & -72.466128 [-72:27:58.06] & 15.402 & 14.310 & -1.092 & -1.794 & e & 246 & ...  \\ 
		76 & 14.049814 [00:56:11.96] & -72.477142 [-72:28:37.71] & 17.319 & 16.113 & -1.206 & -0.913 & e & ... & 1.4  \\ 
		94 & 14.125068 [00:56:30.02] & -72.450499 [-72:27:01.80] & 17.603 & 16.579 & -1.024 & -0.185 & e & ... & ...  \\ 
		95 & 14.048756 [00:56:11.70] & -72.480595 [-72:28:50.14] & 17.921 & 16.989 & -0.931 & -1.161 & e & ... & ...  \\ 
		98 & 14.019112 [00:56:04.59] & -72.453643 [-72:27:13.11] & 17.847 & 16.930 & -0.917 & -0.201 & e & ... & ...  \\ 
		109 & 14.141323 [00:56:33.92] & -72.478912 [-72:28:44.08] & 18.038 & 17.221 & -0.817 & -0.823 & e & 117 & ...  \\ 
		\hline
		47 & 14.042005 [00:56:10.08] & -72.459053 [-72:27:32.59] & 14.481 & 13.178 & -1.303 & -0.436 & a & 45 & ...  \\ 
		49 & 14.044407 [00:56:10.66] & -72.469560 [-72:28:10.42] & 15.184 & 13.678 & -1.507 & ... & a & 47 & ...  \\ 
		53 & 14.039188 [00:56:09.41] & -72.466356 [-72:27:58.88] & 14.452 & 13.199 & -1.253 & -0.432 & a & 49 & ...  \\ 
		66 & 14.073804 [00:56:17.71] & -72.477194 [-72:28:37.90] & 16.594 & 15.132 & -1.462 & -0.284 & a & ... & ...  \\ 
		97 & 14.028371 [00:56:06.81] & -72.476448 [-72:28:35.21] & 15.942 & 14.163 & -1.779 & ... & a & 109 & ...  \\ 
		100 & 14.099138 [00:56:23.79] & -72.483452 [-72:29:00.43] & 16.358 & 14.960 & -1.398 & -0.315 & a & ... & ...  \\ 
		\hline
	\end{tabular}
	\flushleft
	%	\tablenotemark{key letter(s)}
	\tablenotetext{a}{HST magnitudes from~\citet{milone-2018}.}
	\tablenotetext{b}{Color defined as $m_{\rm F336W} - m_{\rm F814W}$.}
	\tablenotetext{c}{H$\alpha$ excess defined as $m_{\rm F656N} - m_{\rm F814W}$.}
	\tablenotetext{d}{H$\alpha$ line shape (emission e, or absorption a).}
	\tablenotetext{e}{$v\sin{i}$ in $\rm km\,s^{-1}$ derived by fitting \texttt{TLUSTY} models to the He I line at 4143.76\,\r{A} and the H$\delta$ line. A 50\,$\rm km\,s^{-1}$ uncertainty on $v\sin{i}$ was estimated for our method.}
	\tablenotetext{f}{Shell parameter defined as the ratio between the average of the blue and red maxima of the H$\alpha$ line and the center of the H$\alpha$ line.}
	%	\tablerefs{reference list}
	%	\tablecomments{Values }
\end{table*}

\begin{table*}
	\centering % Center the table
	\caption{Targets observed with MIKE and their parameters.}
	\label{tab:list-mike}
	\begin{tabular}{cccccccccc}
		\hline
		Target & R.A. (2000.0) [hh:mm:ss] & Decl. (2000.0) [dd:mm:ss] & $m_{\rm F814W}$\tablenotemark{a} & $m_{\rm F336W}$\tablenotemark{a} & Color\tablenotemark{b} & H$\alpha$ excess\tablenotemark{c} & H$\alpha$\tablenotemark{d} & $\vsini$\tablenotemark{e} & shell\tablenotemark{f}\\
		\hline
		52 & 14.050572 [00:56:12.14] & -72.454745 [-72:27:17.08] & 15.232 & 14.003 & -1.229 & -1.419 & e & 161 & ...  \\ 
		56 & 14.073880 [00:56:17.73] & -72.475708 [-72:28:32.55] & 15.509 & 14.373 & -1.135 & -1.611 & e & 220 & ...  \\ 
		58 & 14.059443 [00:56:14.27] & -72.475078 [-72:28:30.28] & 15.726 & 14.689 & -1.037 & -2.085 & e & 270 & ...  \\ 
		68 & 14.106100 [00:56:25.46] & -72.451997 [-72:27:07.19] & 15.366 & 14.213 & -1.152 & -1.727 & e & 55 & 1.2  \\ 
		73 & 14.027701 [00:56:06.65] & -72.460096 [-72:27:36.35] & 14.389 & 13.079 & -1.309 & -1.568 & e & 209 & ...  \\ 
		75 & 14.129855 [00:56:31.17] & -72.466128 [-72:27:58.06] & 15.402 & 14.310 & -1.092 & -1.794 & e & 259 & ...  \\ 
		83 & 14.062130 [00:56:14.91] & -72.479919 [-72:28:47.71] & 15.245 & 14.274 & -0.971 & -1.960 & e & 261 & 1.2  \\ 
		84 & 14.134507 [00:56:32.28] & -72.463979 [-72:27:50.32] & 15.344 & 14.092 & -1.253 & -1.309 & e & 115 & ...  \\ 
		\hline
		51 & 14.113468 [00:56:27.23] & -72.458694 [-72:27:31.30] & 16.310 & 15.011 & -1.299 & -0.303 & a & 24 & ...  \\ 
		69 & 14.060394 [00:56:14.49] & -72.476796 [-72:28:36.47] & 16.370 & 14.967 & -1.402 & -0.322 & a & 48 & ...  \\ 
		88 & 14.128804 [00:56:30.91] & -72.472024 [-72:28:19.29] & 15.728 & 14.271 & -1.457 & -0.333 & a & 34 & ...  \\ 
		100 & 14.099138 [00:56:23.79] & -72.483452 [-72:29:00.43] & 16.358 & 14.960 & -1.398 & -0.315 & a & 67 & ...  \\ 
		\hline
	\end{tabular}
	\flushleft
	%	\tablenotemark{key letter(s)}
	\tablenotetext{a}{HST magnitudes from~\citet{milone-2018}.}
	\tablenotetext{b}{Color defined as $m_{\rm F336W} - m_{\rm F814W}$.}
	\tablenotetext{c}{H$\alpha$ excess defined as $m_{\rm F656N} - m_{\rm F814W}$.}
	\tablenotetext{d}{H$\alpha$ line shape (emission e, or absorption a).}
	\tablenotetext{e}{$v\sin{i}$ in $\rm km\,s^{-1}$ derived by fitting \texttt{TLUSTY} models to five He I lines at 4009.27, 4026.191,  4120.82, 4143.76 and 4387.929\,\r{A}}. A 50\,$\rm km\,s^{-1}$ uncertainty on $v\sin{i}$ was estimated for our method.
	\tablenotetext{f}{Shell parameter defined as the ratio between the average of the blue and red maxima of the H$\alpha$ line and the center of the H$\alpha$ line.}
	%	\tablerefs{reference list}
	%	\tablecomments{Values }
\end{table*}

\section{Observations and data reduction} \label{sec:observations}

Target coordinates and Hubble Space Telescope (HST) magnitudes were extracted from~\citet{milone-2018} and listed in Tables~\ref{tab:list-m2fs}~\&~\ref{tab:list-mike}. Targets observed with the Michigan/Magellan Fiber System~\citep[M2FS,][]{mateo-2012} were selected so that no other star was found within 2$''$ of a target, unless the magnitude of the second star was fainter, with a magnitude difference between the two stars in the F814W filter larger than 1.8.
Targets observed with the Magellan Inamori Kyocera Echelle~\citep[MIKE,][]{bernstein-2003} were selected among the brightest known stars in the cluster. Targets 58, 68, 73, 75 and 100 were observed both with M2FS and MIKE.
%that could be observed from November 27 to 29, 2021.}

MIKE spectra were recorded between November 27 to 29,  2021, with the blue and red arms, covering 3350-9500\,\r{A}. For each target, 2 to 4 exposures of 1200 to 1800\,s were obtained and summed in order to obtain the final spectra.
%\textbf{MIKE} spectra were obtained with the blue and red arms, covering 3350-9500\,\r{A}. \textbf{With MIKE, 2 to 4 exposures of 1200 to 1800s were obtained for each target, and summed in order to obtain the final spectra.
	
With M2FS, spectra were recorded in 2019 and 2022. 
Observations were carried out with two filters:
%This work relied on observations carried out with two M2FS filters: 
a Li-H$\alpha$ filter, and a so-called `Dupree-blue' filter. The Li-H$\alpha$ filter allows us to capture two orders in the $6530$--$6790$\,\r{A} wavelength range for up to 128 targets at once. The Dupree-blue filter, specifically built to target He I lines and never used before this study, was designed to cover $3970$--$4196$\,\r{A} over 4 orders, allowing spectra of up to 64 targets to be recorded simultaneously.

Table~\ref{tab:observations} summarizes the exposure times and number of exposures for each observation night.
All spectra recorded with MIKE and M2FS were doppler shifted to the rest frame assuming a bulk velocity of 180\,$\kms$ for NGC\,330. The resolving powers of MIKE and M2FS  are similar for the considered modes and wavelength ranges (R$\approx$25000 and R$\approx$30000, respectively).

\subsection{M2FS data reduction}

Data were reduced through a Python pipeline largely described in~\citet{walker-2023}, and adapted to meet our needs\footnote{https://github.com/pcristof/M2FS-RS}~\citep{m2fs-rs}.
In this section, we briefly describe the main steps of the reduction.
Raw data consists of images obtained through 4 amplifiers for each of the two detectors. Bias, overscan and dark corrections are applied independently for each amplifier, relying on the Python \texttt{astropy/ccdproc} package~\citep{craig-2017}. For each detector, the 4 images are then joined to generate the full frames used to extract the spectra. Aperture traces were identified on flat images, and used to extract the spectra from science frames. Wavelength calibration was performed by identifying known lines from ThAr exposures obtained right before or between two science exposures.

\begin{figure}
	\centering
	\includegraphics[width=\linewidth]{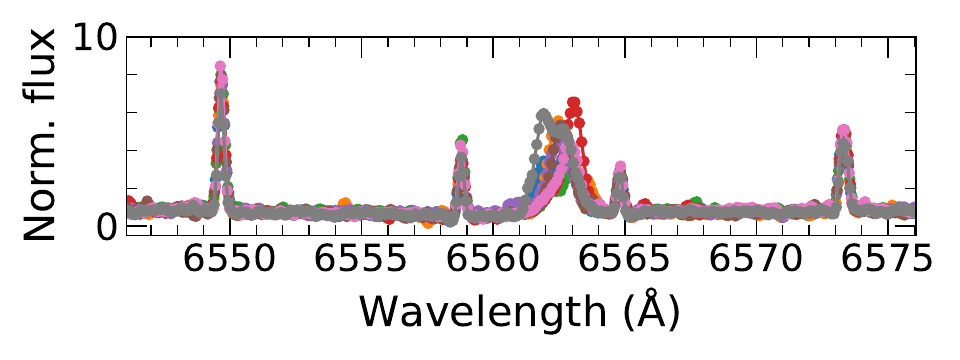}
	\caption{Normalized sky spectra recorded with M2FS. The sharp, thin emission lines are OH lines, while the central line is  H$\alpha$ emission originating from the nebula. The colors correspond to spectra extracted from different sky apertures.}
	\label{fig:m2fs-skies}
\end{figure}

The extracted spectra contain emission sky lines. Several M2FS fibres pointing to the sky were used to extract sky spectra that were combined in order to obtain a template sky spectrum (see Fig.~\ref{fig:m2fs-skies}). For each science spectrum, the intensity of the sky template was adjusted in order to remove the sky emission features. Although this approach efficiently removes OH lines, residual H$\alpha$ emission is visible in the sky spectra and cannot be fully corrected this way (see Sec.~\ref{sec:nebular-emission}).

\subsection{MIKE data reduction}

\begin{figure}
	\centering
	\includegraphics[width=\linewidth]{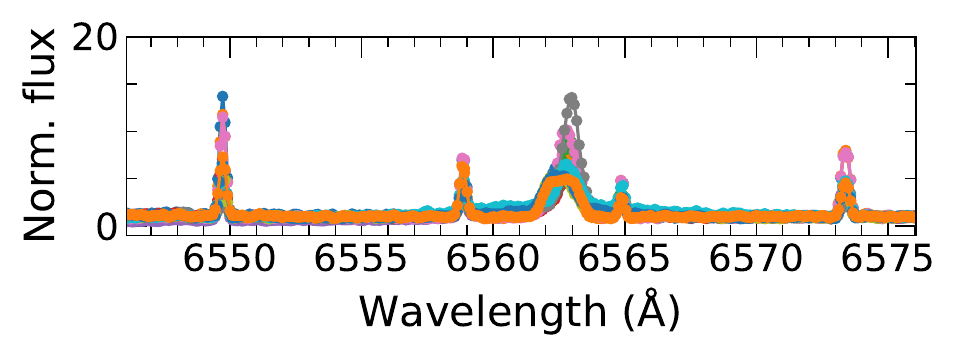}
	\caption{Same as Fig.~\ref{fig:m2fs-skies} for the skies recorded with MIKE. The colors correspond to different sky spectra extracted from the apertures.}
	\label{fig:mike-skies}
\end{figure}

MIKE data were reduced with the Carnegie Python Distribution (CarPy) developed by D. Kelson~\citep{kelson-2000, kelson-2003}. CarPy was used to perform aperture extraction, wavelength calibration, and to normalize the extracted orders. One of the advantages of MIKE over M2FS, as will become clear in Sec.~\ref{sec:nebular-emission}, is the possibility to extract the sky and science spectra from each aperture (see Fig.~\ref{fig:mike-skies}), producing accurate
sky subtractions~\citep{kelson-2003}.


\subsection{Spectral features selection}

H$\alpha$ and He~I lines were the targets of this study in order to assess the presence of Be stars and estimate their projected rotational velocities. With M2FS, the H$\alpha$ line and a He~I line at 4143.76\,\r{A} were extracted from the two filters. With MIKE, several Hydrogen lines from the Balmer series and up to 13 Helium lines were extracted.
Extracted spectra were normalized with a low degree polynomial in the region surrounding those lines. 

\begin{table*}
	\centering
	\caption{Observations used in this work. }
	\label{tab:observations}
	\begin{tabular}{cccccc}
		\hline
		Date & Filter\tablenotemark{a}  & No. exp. & Exp. time (s) & Instrument & Total time (hrs)\\
		\hline
		2019 Nov. 16 & Li-H$\alpha$ & 4 & 2100 & M2FS & 2.3\\
		2022 Sep. 20 & Li-H$\alpha$ & 4 & 1800 & M2FS & 2.0\\
		2022 Sep. 21 & Dupree-blue & 6 & 1800 & M2FS & 3 \\
		2021 Nov. 27 -- 29 & ... & 2 -- 4 &  1200 -- 1800 & MIKE & 0.7 -- 2.0\\
		\hline
	\end{tabular}
	\flushleft
	\tablenotetext{a}{The wavelength coverage of the M2FS filters are 6540--6790\,\r{A} for the Li-H$\alpha$ filter, and 3970-4196\,\r{A} for the Dupree-blue filter. MIKE spectra cover the 3350--9500\,\r{A} wavelength range.}
\end{table*}

\section{Results} \label{sec:results}

\begin{figure}
	\includegraphics[width=\linewidth]{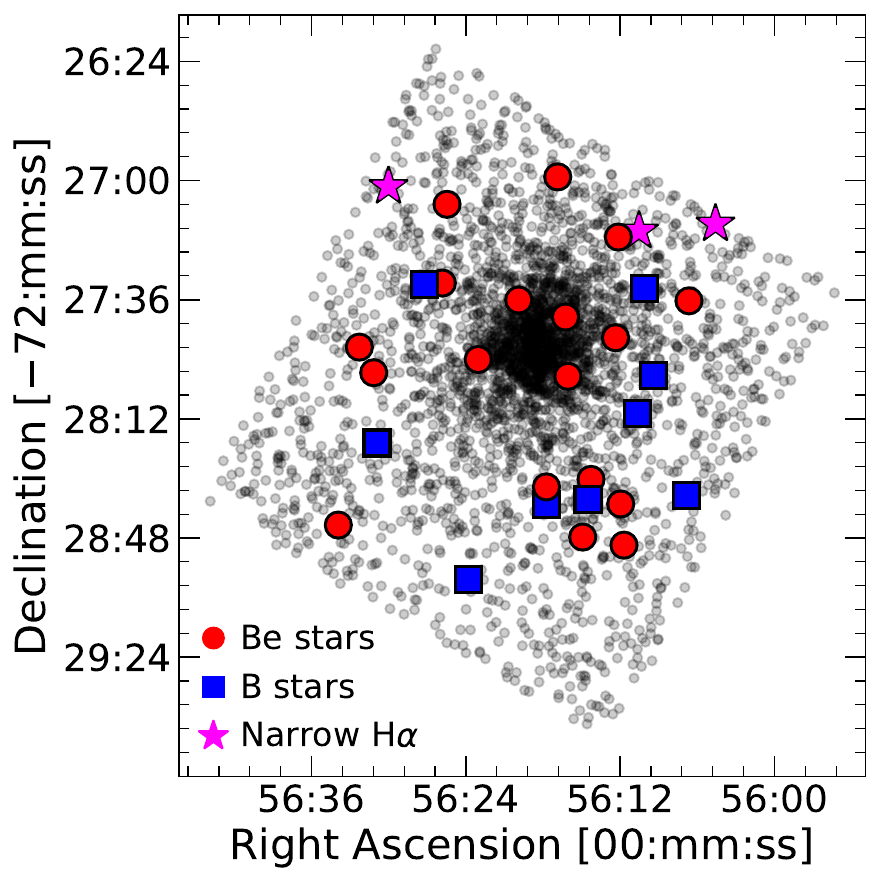}
	\caption{Position of the stars on the sky. Declinations are given in degrees, arc minutes and arc seconds, while right ascensions are in hours, minutes and seconds.  Gray circles mark the position of 2827 targets in the direction of NGC\,330~\citep{milone-2018}. We mark the position of the identified  Be stars (red circles), and B stars (blue squares), and the three narrow H$\alpha$ emitters (fuchsia stars).}
	\label{fig:position-targets}
\end{figure}

M2FS and MIKE high-resolution spectra are used to identify and characterize the H$\alpha$ line, that was visually inspected for all targets, in order to search for emission and or/absorption features, providing information on the presence of a decretion disk around B stars~\citep{rivinius-2013}. Because H$\alpha$ emission lines do not provide direct information on the rotational velocity of the identified Be stars~\citep{dupree-2017}, fits of synthetic spectra were performed on He~I lines to estimate the projected rotational velocity ($v\sin{i}$) of the targets.

\subsection{Nebular emission in NGC~330}
\label{sec:nebular-emission}
Inspection of the spectra recorded with M2FS and MIKE revealed a broad emission feature in the fibers pointing at the sky (see Figs.~\ref{fig:m2fs-skies}~\&~\ref{fig:mike-skies}). 
The shape and intensity of this feature strongly vary among fibers and apertures, and  are  caused by nebular emission. 
With M2FS, sky spectra were not recorded immediately next to each target, but in dedicated fibers pointing in the direction of the nebula. Consequently, accurately removing the nebular features is not possible. Instead, we rely on the OH lines to fit a sky template spectrum and remove an averaged nebular emission feature. Because this correction can be sub-optimal, spectra were inspected both with and without sky subtraction in order to assess the impact of the nebular features on the shape of the stellar H$\alpha$ line. 
The nebular emission significantly contributes to the spectra of faint targets, making it difficult to draw conclusion regarding the core of the stellar H$\alpha$ line. Nebular emission was reported in previous studies of stars in NGC\,330~\citep[e.g., ][]{bodensteiner-2023}. Low-resolution spectroscopy is unlikely to clearly resolve the narrow emission line coming from the nebula.

\subsection{Hydrogen emission features}

Out of the 30 targets observed with M2FS and/or MIKE~(see Fig.~\ref{fig:position-targets}), H$\alpha$ emission features were detected for 21 of them (assumed to indicate the presence of Be stars), and H$\alpha$ broad absorption lines were observed for 9 of them~(see Fig.~\ref{fig:m2fs-1}). Three of those targets have strong but narrower H$\alpha$ emission lines.

The shape of the H$\alpha$ line is expected to provide information on the inclination the targets~\citep{rivinius-2013, sigut-2023}.
For some targets, such as objects 75 and 58, the `wine-bottle'  shape of the H$\alpha$ line is typically associated with low inclination angles (with a star viewed almost pole-on).
 Other targets (e.g., 18, 64, 68, 83) show signs of deep core reversal with various depths, expected for stars viewed with different inclination angles~\citep{rivinius-2013, sigut-2023}. 
 The inaccurately corrected nebular emission in the M2FS spectra renders the precise study of the H$\alpha$ line core difficult for some targets. 
Previous studies~\citep{hanuschik-1996,sigut-2023} have reported a link between the shell parameter (i.e. the ratio between the average of the maxima on each side of the H$\alpha$ line center and the H$\alpha$ line center) and the inclination of the star.
In particular, ~\citet{sigut-2023} found that shell parameters above 1.5 tend to indicate inclinations above 75$^\circ$.
Shell parameters derived from the M2FS and MIKE spectra are reported in Tables~\ref{tab:list-m2fs} and~\ref{tab:list-mike}, and were found to be above 1.5 for only 2 targets (18 and 64) suggesting that most of the other stars in our sample have inclinations below 75$^{\circ}$ (here, $i=0^{\circ}$ corresponds to a star seen pole-on and $i=90^{\circ}$ implies that the rotation axis is perpendicular to the line of sight).
Large asymmetries were observed in the H$\alpha$ lines for several targets (e.g. targets 7, 64, 68, 73, 95). We note that if the observed targets were interacting binaries, the complexity of the circumstellar disk is expected to impact the shape of the H$\alpha$ line~\citep{zorec-2023}.

\begin{figure}
	\centering
	\includegraphics[width=0.85\linewidth, trim={0.3cm, 0.3cm, 0.3cm, 0.25cm}, clip]{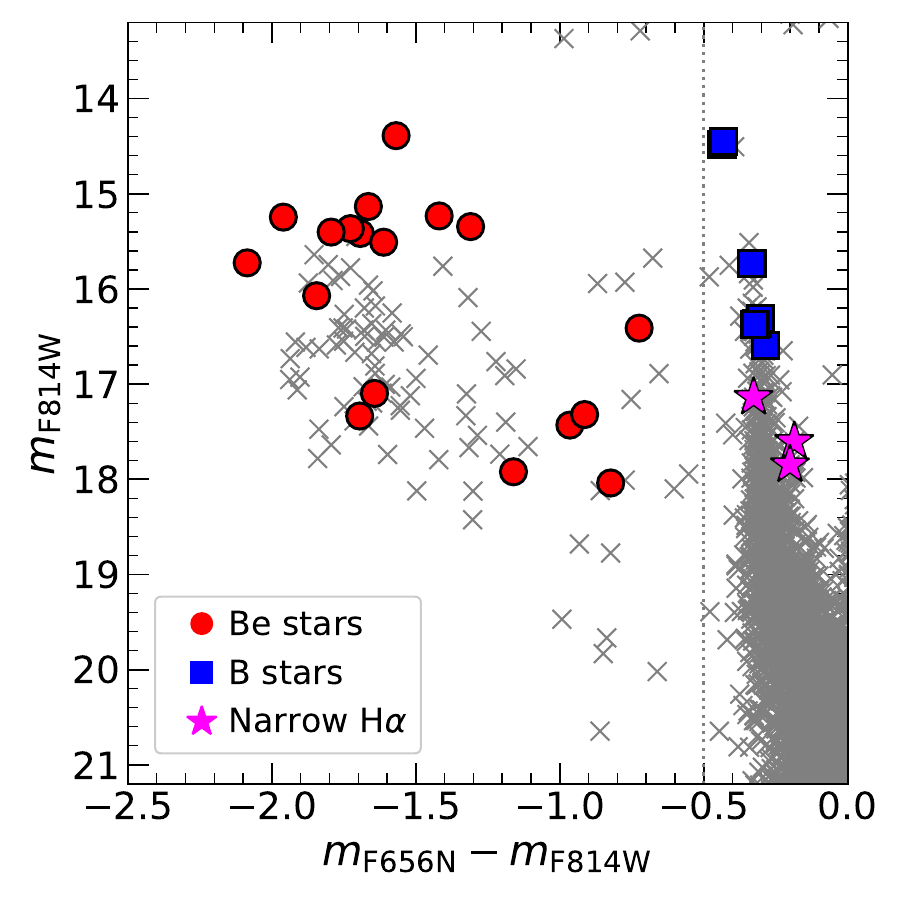}
	\caption{Comparison between the F814W magnitude and relative excess in the H$\alpha$ F656N filter. Gray crosses mark the position of 2827 stars in NGC330~\citep{milone-2018}.  Red circles and blue squares mark the position of Be stars and B stars, respectively. Fuchsia stars mark the position of 3 `narrow H$\alpha$ emitters' for which narrow H$\alpha$ and H$\delta$ emission lines were observed. The dashed gray line shows the fiducial boundary at $m_{\rm F656N}-m_{\rm F814W}=-0.5$ separating the population of H$\alpha$ emitters and absorbers from HST photometry~\citep{milone-2018}. }
	\label{fig:cmd-2}
\end{figure}

The photometric data acquired through the Hubble F656N H$\alpha$ filter~(see Fig.~\ref{fig:cmd-2}) reveal excesses consistent with the identified broad H$\alpha$ emission features. 
Among the stars for which H$\alpha$ emission was detected, a few of them (e.g. objects 2, 76, 95, 109, see Fig.~\ref{fig:m2fs-1}) show significantly shallower emission than others, and correspond to the stars whose excess in the H$\alpha$ filter ($m_{\rm F656N} - m_{\rm F814W}$) are found between -0.5 and -1.0 (see Fig.~\ref{fig:cmd-2}).

Emission in other Hydrogen lines from the Balmer series is expected to originate from regions closer to the central star than H$\alpha$ lines. For several Be stars, H$\delta$ emission features were observed at the center of H$\delta$ photospheric absorption. The H$\delta$ emission appears sharp for several targets whose $v\sin{i}$ are  lower than 100\,$\rm km\,s^{-1}$ (see Sec.~\ref{sec:vsinis}).
Narrow deep absorption features are also observed  at the center for the H$\delta$ line for targets 64 and 18 that have deep H$\alpha$ core reversal, and for target 83, which could result from absorption caused by the decretion disk at large inclination angles. Such emission features are also observed in the spectra recorded with MIKE for several hydrogen lines of the Balmer series; in particular for target 68 and 84 (see Fig.~\ref{fig:full-mike}).

The analysis of the MIKE spectra revealed additional features, including up to 8 Fe II emission lines (see Table~\ref{tab:linelist}). 
	For targets 58, 68, 73, 83 and 84, two of those Fe II emission lines appear on the red side of the helium lines at 4922\,\r{A} and 5015\,\r{A}, leading to line shapes that resemble a P~Cygni profile.
	 For several of those targets, the cores of the helium lines at 5875, 6678 and 7065\,\r{A} are also shallower than predicted by the models, often asymmetric, or reveal helium emission lines. Those features could indicate the presence of winds.

\begin{table}
	\caption{He I and Fe II lines identified in the MIKE spectra.}
	\label{tab:linelist}
	\begin{tabular}{ccc}
		\hline
		Wavelength (\r{A}) & Species & Ref.\tablenotemark{a} \\
		\hline
		3634.23 & He I & NIST \\ 
		3634.24 & He I & NIST \\ 
		3651.99 & He I & NIST \\ 
		3705.01 & He I & NIST \\ 
		3819.79 & He I & NIST \\ 
		3867.48 & He I & NIST \\ 
		3871.79 & He I & NIST \\ 
		3964.73 & He I & NIST \\ 
		4009.27 & He I & NIST \\ 
		4026.19 & He I & NIST \\ 
		4120.82 & He I & NIST \\ 
		4143.76 & He I & NIST \\ 
		4169.18 & He I & NIST \\ 
		4387.93 & He I & NIST \\ 
		4437.55 & He I & NIST \\ 
		4471.48 & He I & NIST \\ 
		4713.31 & He I & NIST \\ 
		4921.93 & He I & NIST \\ 
		5015.68 & He I & NIST \\ 
		5047.74 & He I & NIST \\ 
		5875.61 & He I & NIST \\ 
		6678.15 & He I & NIST \\ 
		7065.18 & He I & NIST \\ 
		\hline
		4549.20 & Fe II & NIST \\ 
		4923.92 & Fe II & NIST \\ 
		5018.44 & Fe II & NIST \\ 
		5169.03 & Fe II & NIST \\ 
		5316.61 & Fe II & NIST \\ 
		6318.12 & Fe II & K13 \\ 
		6347.55 & Fe II & NIST \\ 
		6371.72 & Fe II & NIST \\ 
		\hline
	\end{tabular}
	\tablenotetext{a}{Reference used for the identification of the spectral line. \\ NIST: NIST database version 5.11~\citep{nist-asd-v511}. \\ K13: R. Kurucz calculations, 2013, available at \texttt{http://kurucz.harvard.edu/atoms/2601/gfemq2601.pos} and accessed through the VALD database.}
\end{table}

\subsection{\texorpdfstring{Narrow H$\alpha$ emitters}{Lg}}

The spectra of a few targets, such as 61, 94 or 98, display intense and narrow H$\alpha$ emission lines, whose interpretation is limited by the presence of nebular emission (see Fig.~\ref{fig:m2fs-1}). Clear photospheric H$\delta$ absorption lines are observed for the three targets, and a narrow emission feature is visible at the center of these absorption lines. We refer to those three targets as `narrow H$\alpha$ emitters' in the rest of the paper.

The full width at half maximum (FWHM) of the H$\delta$ and H$\alpha$ emission features are on the order of  $\sim$50\,$\rm km\,s^{-1}$ for the three narrow H$\alpha$ emitters, while the equivalent widths of the emission component of the H$\alpha$ lines are significantly larger than that of H$\delta$ lines. This is consistent with the spectra expected for Be stars, with H$\delta$ lines originating from regions in the disk closer to the central star and hence weaker. Larger equivalent widths for H$\alpha$ lines than for other hydrogen lines are found for all Be stars in our sample.

\begin{figure}
		\centering
	\includegraphics[width=.9\columnwidth]{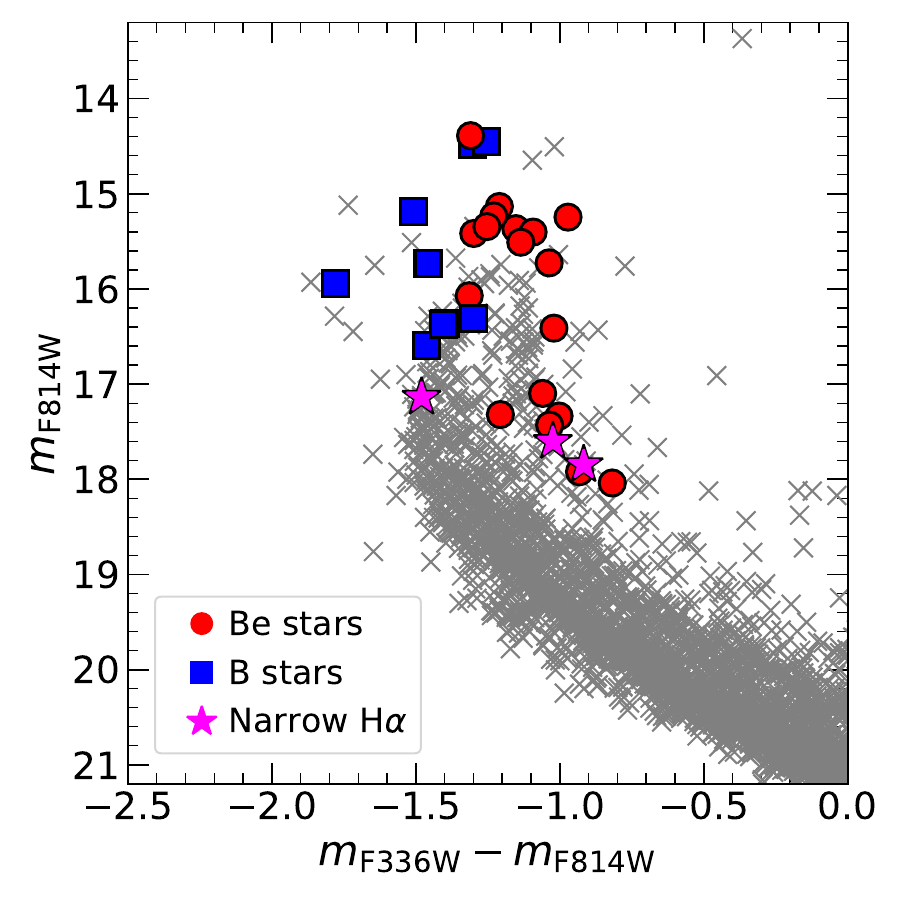}
	\caption{Color-magnitude diagram for NGC330 targets. Gray crosses mark the position of 2827 stars in NGC330~\citep{milone-2018}. 
		Red circles and blue squares mark the position of Be stars and B stars, respectively. Fuchsia stars mark the position of the 3 `narrow H$\alpha$ emitters' for which narrow H$\alpha$ and H$\delta$ emission lines were observed.}
	\label{fig:cmd}
\end{figure}

\subsection{Temporal evolution of \texorpdfstring{H$\alpha$}{} profiles}

\begin{figure*}
	\includegraphics[width=0.99\linewidth]{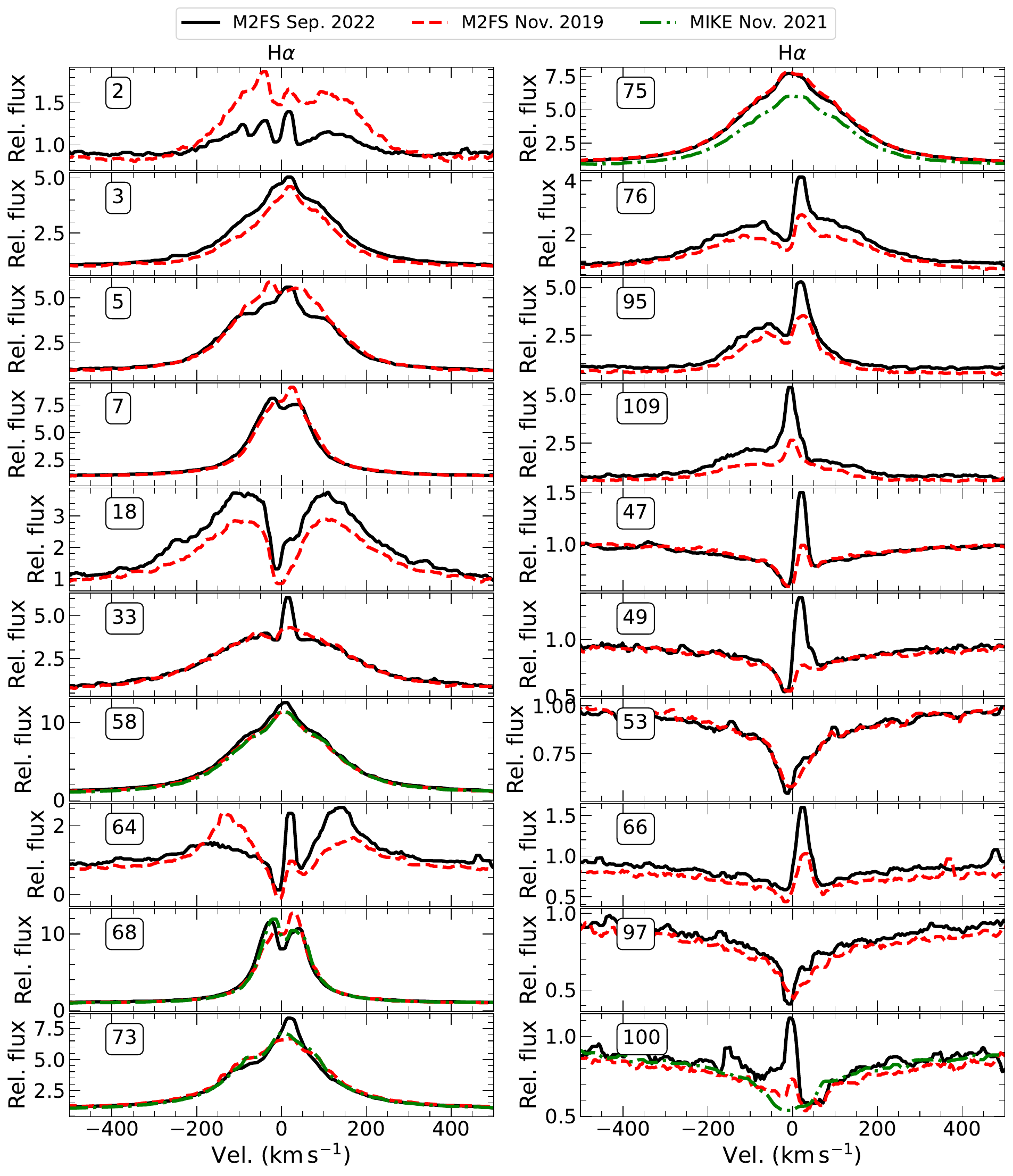}
	\centering
	\caption{Comparison between sky-corrected spectra recorded in 2019 (red broken line) and 2022 (solid black line) with M2FS, and in 2021 with MIKE (dotted-dashed green line).}
	\label{fig:m2fs-targets-comparison}
\end{figure*}

The spectra recorded with M2FS in 2019 and 2022, and with MIKE in 2021 reveal variations of the H$\alpha$ emission line shape (see Fig.~\ref{fig:m2fs-targets-comparison}). For some targets, such as 2, 18 or 75, variations in the strength of the emission lines are observed. For other targets, such as 64 or 68, changes in the asymmetry of the line shapes are visible. For target 64, the H$\alpha$ line exhibits two main peaks, with the blue side of the line being more intense than the red side in 2019. In 2022 the asymmetry is reversed: the red side is stronger than the blue side.
%, a similar profile is observed, but the red side is then more intense than the blue side of the H$\alpha$ emission line. 
A somewhat similar evolution is observed for target 68 between 2019 and 2021/2022.

The evolution of the H$\alpha$ line shape tracks movement in the emitting regions of the decretion disk. Variations in the strength of the H$\alpha$ line, in particular, have been proposed to arise from fluctuations in the extent of the emission region~\citep{vinicius-2006}. 
Slowly revolving perturbation patterns, sometimes referred to as `one-armed' oscillations~\citep{okazaki-1991, okazaki-1997, papaloizou-2006}, have been proposed to explain asymmetric hydrogen line profiles and their evolution, in which excess emission in the blue or red side of a spectral line is attributed to enhanced temperature and density in the approaching or receding side of the disk. This is further supported by the observation of similar asymmetries in the H$\beta$, H$\gamma$,  and H$\delta$ lines recorded with MIKE for targets 52, 68, 73 and 84 (see Figs.~\ref{fig:hydrogen-lines}).

\subsection{Rotational velocities}
\label{sec:vsinis}
Because the rotational velocities of the targeted stars are expected to be large, we assumed that rotation is the main source of broadening in the spectra.
Following the results of~\citet{bodensteiner-2023}, we derived $\vsini$ measurements by fitting synthetic models taken from the \texttt{TLUSTY} BSTAR2006 grid~\citep{hubeny-1995, lanz-2007} to the spectra. For spectra recorded with M2FS, the fit is performed on the He I line at 4143.76\,\r{A} and the H$\delta$ line making sure to ignore the central emission feature. For spectra recorded with MIKE, we rely on 5 of the best modeled He I lines at 4009.27, 4026.191,  4120.82, 4143.76 and 4387.929\,\r{A}, allowing us to improve the fits to the data. 
	
For each spectrum, a least-squares fit is performed to obtain the best fitting \texttt{TLUSTY} model. Specifically, rotational velocity is included in the model by convolving the synthetic spectra with a rotation profile. A fit is obtained for each combination of effective temperature ($T_{\rm eff}$) and log surface gravity ($\log{g}$) assuming a metallicity $Z = 0.1\,Z_{\sun}$~\citep{piatti-2019}, leading to computation of a grid of $\chi^2$ values and associated $\vsini$. The fit leading to the lowest $\chi^2$ is then considered the best match, and used to obtain $\vsini$.
	This approach provides a very rough estimate of $T_{\rm eff}$ and $\log{g}$, given that the fit is obtained from a limited number of spectral features. Several tests have allowed us to confirm that varying $T_{\rm eff}$ by up a few thousands of kelvins or $\log{g}$ by 0.5\,dex had limited impact on the modeled line shape and the derived $\vsini$, with typical differences of less than 50\,$\kms$. This impact is, however, larger than the formal uncertainties computed from the spectral fit. Given the degeneracy and the large uncertainties on the atmospheric parameters, conservative error bars of $\pm$50\,$\kms$ on the $\vsini$ measurements were adopted, to account for some of the systematic errors of the method.

\begin{figure}[ht]
	\centering
	\includegraphics[width=.95\linewidth]{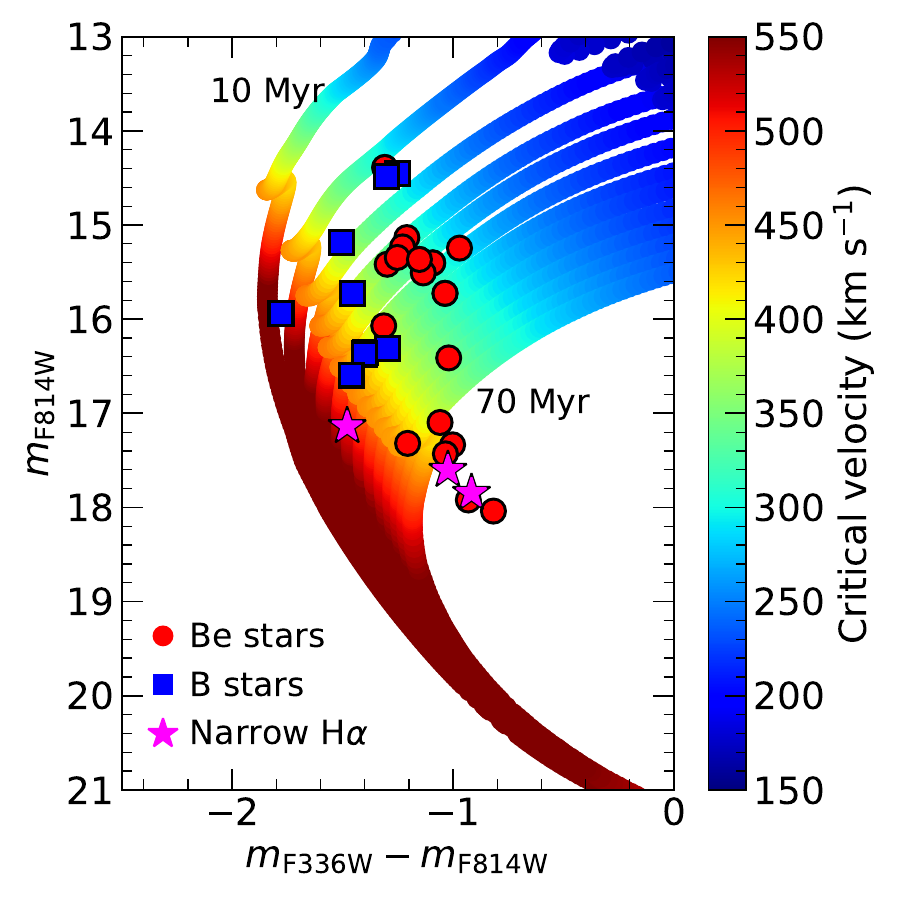}
	\centering
	\caption{Breakup velocities for MIST isochrones  with ages ranging from 10 to 70~Myr and $\rm [Fe/H]=-0.85$\,dex. Red circles and blue squares mark the position of Be stars and B stars, respectively. Fuchsia stars mark the position of the 3 `narrow H$\alpha$ emitters' for which narrow H$\alpha$ and H$\delta$ emission lines were observed.}
	\label{fig:breakup-vel}
\end{figure}

\begin{figure}
	\includegraphics[width=\columnwidth]{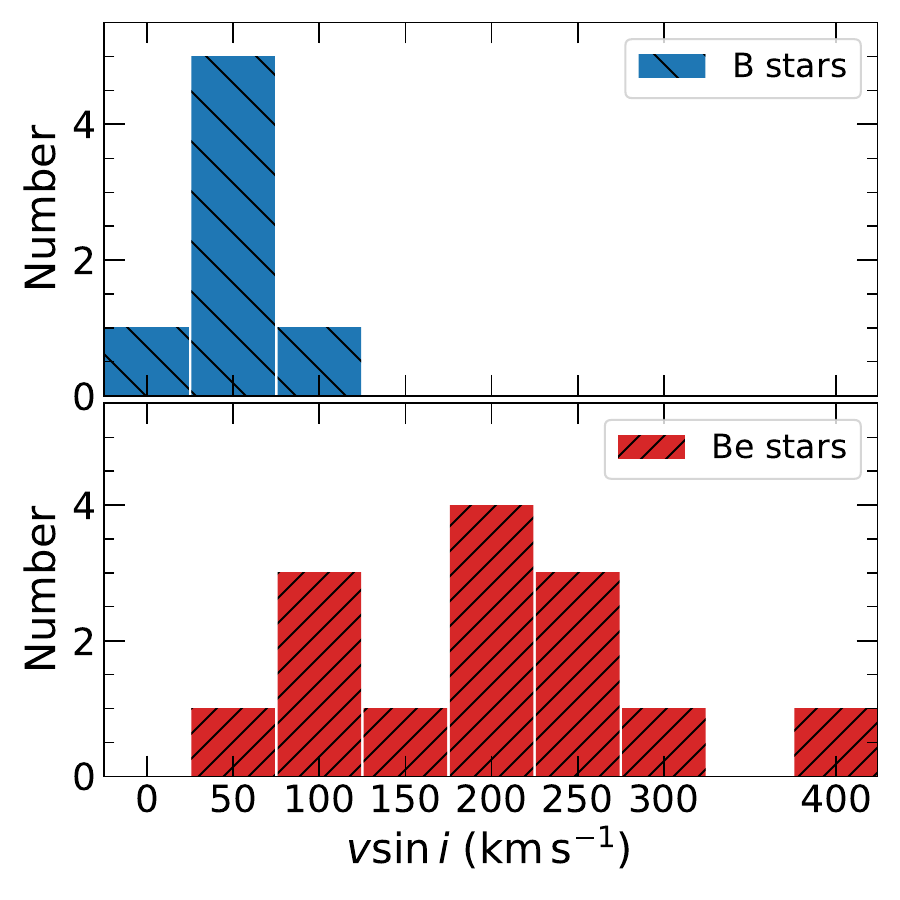}
	\caption{Distribution of the rotation velocities for the targets in our sample.}
	\label{fig:vsini-distrib}
\end{figure}

\begin{figure}
	\includegraphics[width=\columnwidth]{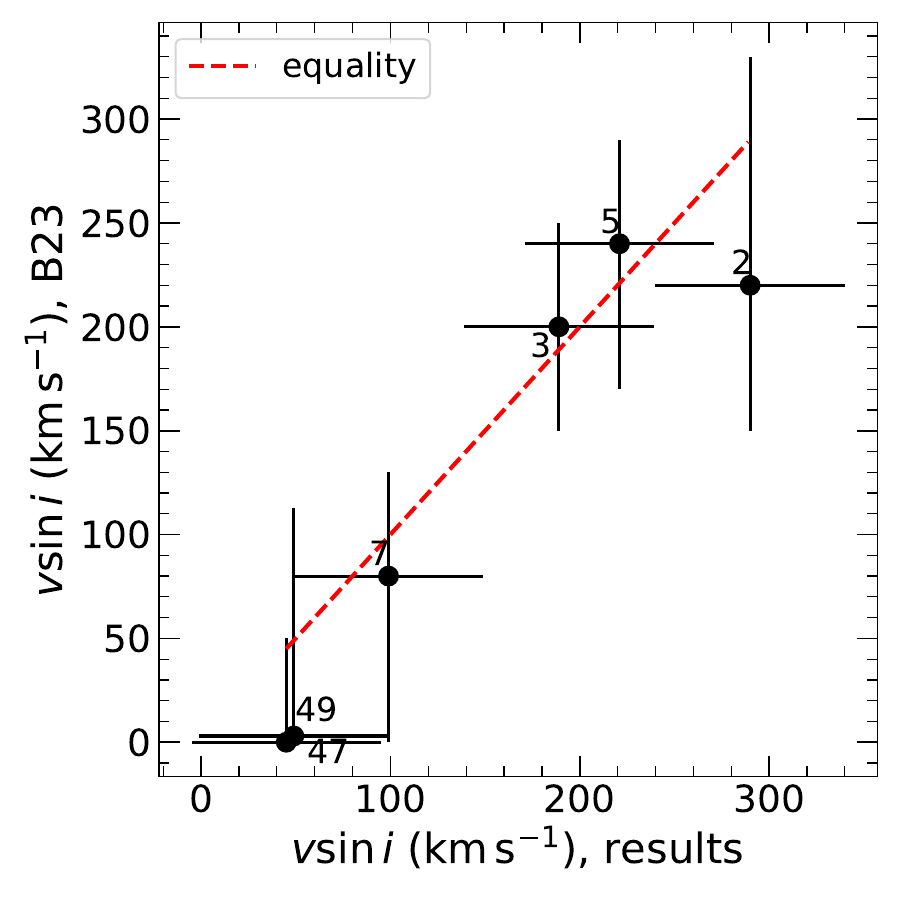}
	\caption{Comparison between the derived $v\sin{i}$ values and those of~\citet[][B23]{bodensteiner-2023} for the 6 targets common to  both studies. The red broken line marks the equality.}
	\label{fig:boden-vsinis}
\end{figure}

\begin{figure}
	\includegraphics[width=\columnwidth]{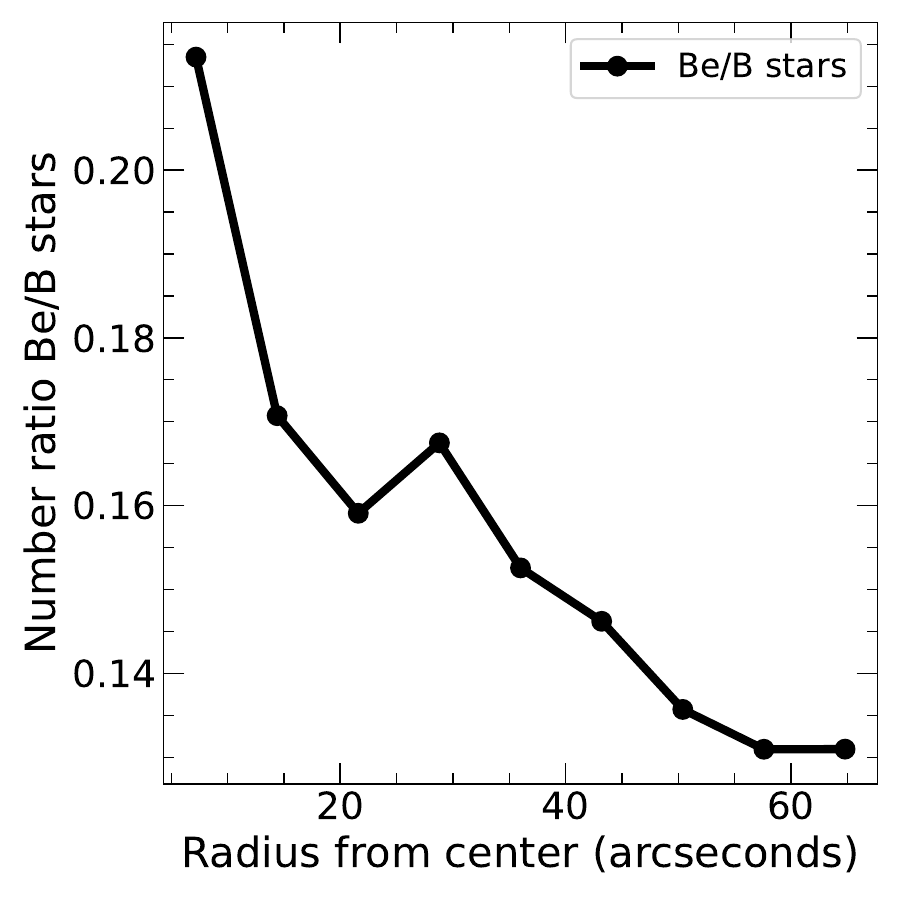}
	\caption{Number ratio of $N_{\rm Be} / N_{\rm B}$, with $N_{\rm Be}$ and $N_{\rm B}$ the number of Be and B stars found within a given radius from the cluster center, respectively. B and Be stars numbers were estimated from excess in the F656N HST filter~\citep{milone-2018}.}
	\label{fig:be-b-ratios}
\end{figure}

For targets identified as B stars, our values of $\vsini$ range from about 30 to $100\,\kms$. Although this contrasts with previous results suggesting a larger range of rotational velocities for B stars in the cluster~\citep{bodensteiner-2023}, 
 	our targets are in the turnoff region of the CMD, and our results are consistent with the lower $\vsini$ found by~\citet{bodensteiner-2023} in this part of the diagram.

For H$\alpha$ emitters, however, our $\vsini$ estimates span a wide range values from $\sim50$ to $\sim$\,$400$\,$\kms$.
We note that the 3 targets 7, 68 and 84 are the only stars with broad H$\alpha$  emission lines in our sample with $\vsini<100\,\kms$. The $v\sin{i}$ estimates obtained for most stars in our sample are consistent with the theoretical critical velocities computed with the MIST models~\citep{paxton-2011, paxton-2013, paxton-2015, choi-2016, dotter-2016} for ages ranging between 10 to 70 Myr and a metallicity $\rm [Fe/H]=-0.85$\,dex (see Fig.~\ref{fig:breakup-vel}). Those models predict critical velocities around 300 and 400\,$\rm km\,s^{-1}$ for most of the Be stars in our sample.

The distribution of $v\sin{i}$ Be stars is centered on $\sim$200\,$\rm km\,s^{-1}$ (see Fig.~\ref{fig:vsini-distrib}). Assuming no selection bias, the distribution is consistent with a range of rotational velocities within the sample of Be stars, given that a random distribution of inclinations would favor high $\sin{i}$ values. The shell parameters obtained for the sample of stars, however, suggest that most of the targets have inclinations below 75$^{\circ}$.

Five stars were observed both with M2FS and MIKE: targets 58, 68, 73, 75 and 100.  For target 68, the $\vsini$ estimates derived from M2FS and MIKE spectra differ by less  than 1\,$\kms$. 
For targets 58, 73 and 75, the estimates agree within $\sim$\,40\,$\kms$, less than our estimated error bar of 50\,$\kms$.
The spectra recorded with M2FS for target 100 did not allow us to derive an estimate of $\vsini$. 
A comparison between the obtained $v\sin{i}$ and those reported by~\citet{bodensteiner-2023} is shown in Fig.~\ref{fig:boden-vsinis}. For the 6 targets included in both studies, the estimates agree within the error bars.
The distribution of $v\sin{i}$ estimates obtained in this study are presented in Fig.~\ref{fig:vsini-distrib}, while the values are reported in Tables~\ref{tab:list-m2fs} and~\ref{tab:list-mike}.

\section{Discussion and conclusions}
\label{sec:discussion}

Most of the H$\alpha$ emitters, assumed to be Be stars, are found in the red side of the split eMSTO, while the other targets fall within the blue branch of the split eMSTO. 
Those results confirm the detection of Be stars from photometric measurements~\citep{milone-2018} suggesting that the red branch of the split eMSTO is populated in the vast majority by Be stars. A clear difference is found between the average $\vsini$ of the Be stellar population ($<\!\vsini\!>\approx200$\,$\kms$) and that made up of stars with no H$\alpha$ emission ($<\!\vsini\!>\approx50$\,$\kms$). The $v\sin{i}$ estimates are consistent with the expected critical velocities computed from MIST models, predicting breakup velocities around 300-400\,$\kms$ for the Be stars in our sample. 
The results also confirm the previous measurements derived from low-resolution data~\citep{bodensteiner-2023}, with excellent agreement between the obtained $v\sin{i}$ for the 6 targets common to both samples.

Computed shell parameters suggest that most stars in the sample have inclinations below $75^{\circ}$, effectively leading to an unexpectedly low number of high-inclination stars. 
It should be noted, however, that~\citet{sigut-2023} also reports larger inclinations and shell parameters below 1.5 for some stars they have analyzed.
	Our $v\sin{i}$ estimates suggest, nonetheless, a range of rotational velocities in the population of Be stars. This is further supported by the observed H$\alpha$ line shapes,  with several targets showing no clear sign of core reversal or asymmetry, in spite of large $v\sin{i}$ estimates (e.g. targets 58, 75), while the `wine-bottle' shape of the lines is believed to arise from targets seen at low inclinations~\citep{rivinius-2013}.

The observation of H$\delta$ emission features supports the presence of a decretion disk. H$\delta$ emission is expected to originate from regions closer to the central star~\citep{klement-2017, zorec-2023}, and are not typical signatures of nebular emission (as confirmed by their absence from sky spectra). 
	Two peculiar cases stand out in our data. First, for the three narrow H$\alpha$ emitters,  H$\delta$ emission features were observed (targets, 61, 94, 98, see Fig.~\ref{fig:m2fs-1}). Although part of the H$\alpha$ emission can originate from the nebula, the presence of photospheric H$\delta$ absorption lines, and narrow H$\delta$ emission in their core, suggest the presence of a Be star. No H$\alpha$ excess is detected in the Hubble F656N H$\alpha$ filter for those targets, most likely because of the narrow width of the observed lines. Thus, those targets could easily be inaccurately classified from HST narrow band photometry, and the Be phenomenon actually more widespread. 
	Secondly, a narrow absorption feature is observed at the center of the reversed core of the H$\delta$ line for target 83. This feature could be the result of absorption from the decretion disk if the star is seen at a large inclination angle, as described for H$\alpha$ lines in shell stars~\citep[e.g.,][]{rivinius-2013, sigut-2023}. This contrasts with the shell parameter of target 83, however, estimated to 1.2, which suggests an inclination lower than 70$^\circ$~\citep{sigut-2023}. From MIKE data, absorption features for target 83 are observed at the center of the H$\alpha$, H$\beta$, H$\gamma$ and H$\delta$, which further supports absorption from the decretion disk (see Fig.~\ref{fig:hydrogen-lines}). 
	Those results suggest that the inclination of some of the stars is difficult to estimate from shell parameters.
 Additional high resolution spectra recorded with, e.g., MIKE targeting H$\alpha$, H$\beta$, H$\gamma$ and H$\delta$, along with accurate modeling of the emission lines in Be stars~\citep[see, e.g.,][]{sigut-2023}, would allow us to provide better constraints on the inclination angle of the observed targets, and to accurately classify the narrow H$\alpha$ emitters identified in this work.

With both M2FS and MIKE spectra,  H$\alpha$ nebular emission was observed. The nebular emission is successfully corrected by the MIKE reduction pipeline, extracting the sky and science spectra from a single aperture, but is impossible to accurately correct for observations collected with M2FS with the setup currently used.
The narrow width of the nebular emission (of a few dozens of pixels only) is unlikely to significantly impact the results of photometric studies observing the broad and strong H$\alpha$ emission of bright targets. However, its contribution is more significant for the measurements of faint stars, in particular if they are slowly rotating and have narrower stellar H$\alpha$ features suggesting that Be stars may be more frequent than presently thought.
Future observations in clusters with nebular emission envisioned with fiber systems should attempt to measure spectra on sky in regions immediately adjacent to the targets positions in order to provide a better correction of those features. 
	We find no B star close to the cluster core (see Fig.~\ref{fig:position-targets}). Although our selection criteria for M2FS introduce a bias towards choosing isolated targets outside of the crowded cluster core, this is in line with the previous results~\citep{bodensteiner-2020} suggesting that the ratio of Be stars in the core is higher than in the outskirts of the cluster. Those results are also consistent with the number of B and Be stars estimated from HST photometry inside and outside of the cluster core (see Fig.~\ref{fig:be-b-ratios}). The narrow H$\alpha$ emitters were detected outside of the cluster core, however, which could indicate that the number of Be stars in the outskirts of the cluster is also underestimated. It should be noted that the bias towards bright targets is likely to lead to higher Be detection rates in our sample than in other studies, given the larger overall content in Be stars in the photometric data published by~\citet{milone-2018} 
		for brighter targets.
Further high-resolution spectroscopic observations of NGC\,330 will help to better characterize the observed stars by allowing us to probe the the core of the emission lines which is necessary to better constrain the properties of the targets, and better understand the multiple populations observed in the CMD. 
In particular, future observations carried out with high resolution spectrographs could be used to search for features indicative of binary systems.

%% Also note that the akcnowlodgment environment does not support long amounts of text. If you have a lot of people and institutions to acknowledge, do not use this command. Instead, create a new \section{Acknowledgments}.
\begin{acknowledgments}
	
	This research has made use of NASA’s Astrophysics Data System Bibliographic Services and of the data products from 2MASS, which is a joint project of the University of Massachusetts and IPAC/Caltech, funded by NASA and the NSF.

	This work has made use of the VALD database, operated at Uppsala University, the Institute of Astronomy RAS in Moscow, and the University of Vienna.

	This work has received funding from ``PRIN 2022 2022MMEB9W - {\it Understanding the formation of globular clusters with their multiple stellar generations}'' (PI Anna F.\,Marino).
	
    We thank Christian Johnson for preparing the fibre configurations for
	M2FS observations and Morgan MacLeod for fruitful discussions regarding the
	content of this paper. We thank Aaron Sigut for discussions regarding the modeling of line profiles.

\end{acknowledgments}


\vspace{5mm}
\facility{Magellan: Clay (M2FS, MIKE).}


\software{Astropy \citep{astropy:2013, astropy:2018, astropy:2022} / ccdproc \citep{astropy-2013, craig-2017},
				CarPy~\citep{kelson-2000, kelson-2003},
				Numba~\citep{lam-2015}.
%          Cloudy \citep{2013RMxAA..49..137F}, 
%          Source Extractor \citep{1996A&AS..117..393B}
          }



\appendix
\restartappendixnumbering
\section{Best fits to the data}

\begin{figure*}
%	\figurenum{A1}
	\plotone{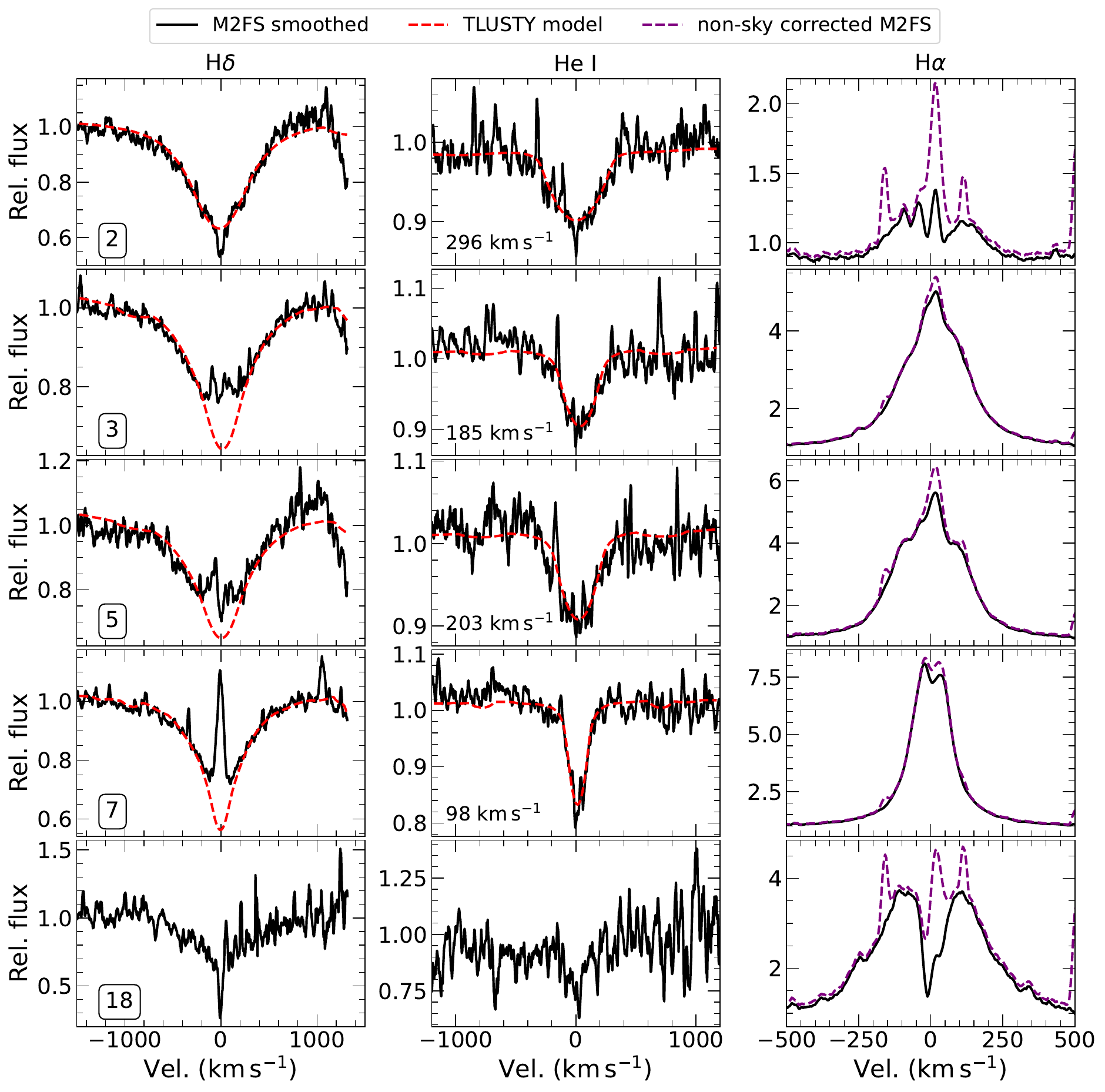}
	\caption{H$\delta$, He I line at 4143.76\,\r{A} and H$\alpha$ line profiles observed with M2FS in 2022 for 5 targets in NGC\,330. The H$\alpha$ line is presented before and after sky correction. The best obtained fit of TLUSTY models to the H$\delta$ and He I lines are shown in dashed red, when available. Projected rotational velocity estimated from the fits are indicated when available.}
	\label{fig:m2fs-1}
\end{figure*}

\begin{figure*}
%	\figurenum{A2}
	\plotone{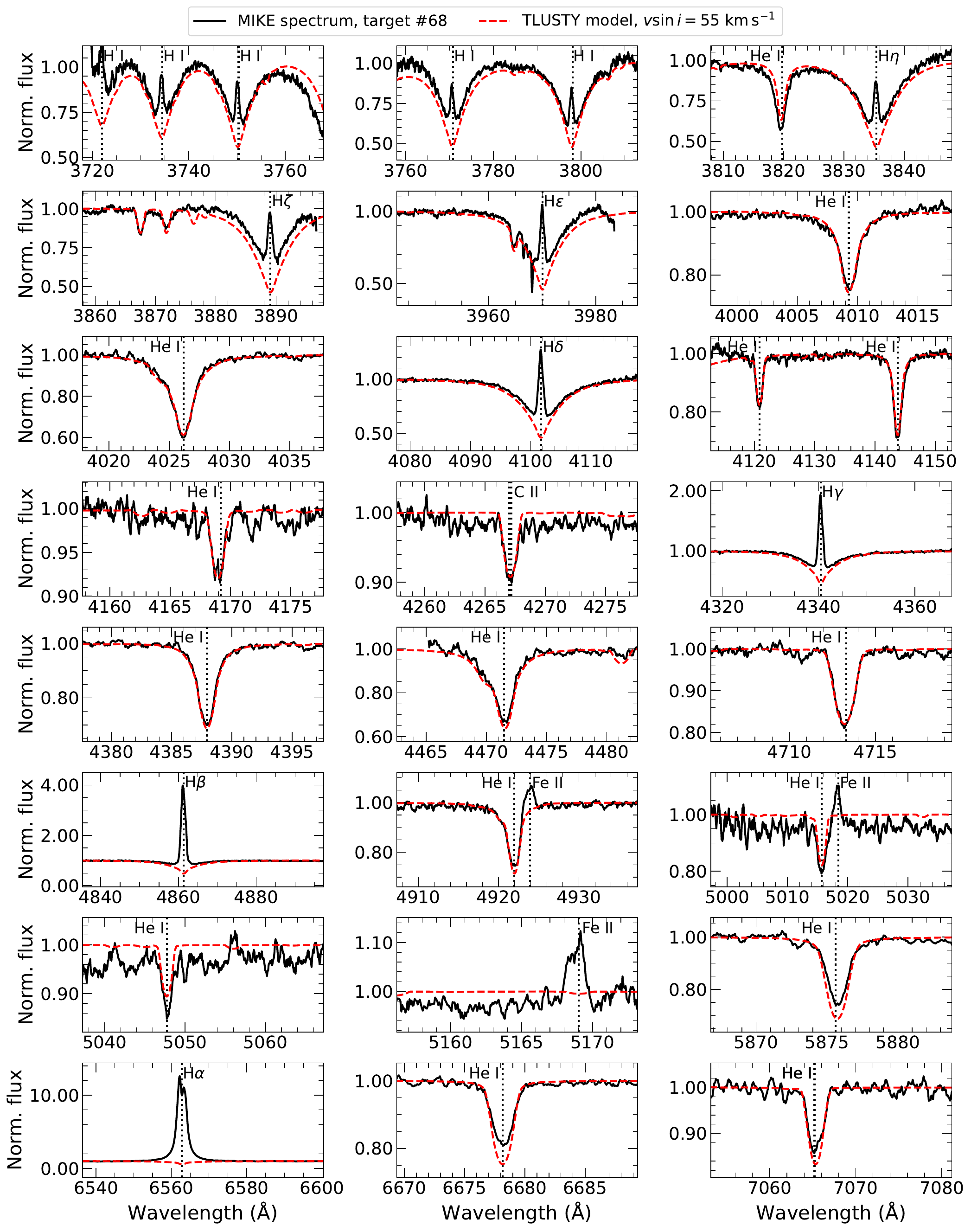}
	\caption{Hydrogen, helium, carbon and iron lines identified in the MIKE spectrum of target 68. The best fit of TLUSTY models is shown in dashed red. }
	\label{fig:full-mike}
\end{figure*}

For the targets studied in this paper, Fig.~\ref{fig:m2fs-1} presents the spectra recorded with M2FS in 2022, and Fig.~\ref{fig:full-mike} presents the spectra recorded with MIKE in 2021. On each figure, we show the best fit of the TLUSTY model to the He I and H$\delta$ lines.

\figsetstart
\figsetnum{A1}
\figsettitle{H$\delta$, He I line at 4143.76\,\r{A} and H$\alpha$ line profiles observed with M2FS in 2022.}

\figsetgrpstart
\figsetgrpnum{A1.1}
\figsetgrptitle{Targets 2 to 18}
\figsetplot{m2fs-2}
\figsetgrpnote{H$\delta$, He I line at 4143.76\,\r{A} and H$\alpha$ line profiles observed with M2FS in 2022. The H$\alpha$ line is presented before and after sky correction. The best obtained fit of TLUSTY models to the H$\delta$ and He I lines are shown in dashed red.}
\figsetgrpend

\figsetgrpstart
\figsetgrpnum{A1.2}
\figsetgrptitle{Targets 33 to 73}
\figsetplot{m2fs-3}
\figsetgrpnote{H$\delta$, He I line at 4143.76\,\r{A} and H$\alpha$ line profiles observed with M2FS in 2022. The H$\alpha$ line is presented before and after sky correction. The best obtained fit of TLUSTY models to the H$\delta$ and He I lines are shown in dashed red.}
\figsetgrpend

\figsetgrpstart
\figsetgrpnum{A1.3}
\figsetgrptitle{Targets 2 to 18}
\figsetplot{m2fs-4}
\figsetgrpnote{H$\delta$, He I line at 4143.76\,\r{A} and H$\alpha$ line profiles observed with M2FS in 2022. The H$\alpha$ line is presented before and after sky correction. The best obtained fit of TLUSTY models to the H$\delta$ and He I lines are shown in dashed red.}
\figsetgrpend

\figsetgrpstart
\figsetgrpnum{A1.4}
\figsetgrptitle{Targets 2 to 18}
\figsetplot{m2fs-5}
\figsetgrpnote{H$\delta$, He I line at 4143.76\,\r{A} and H$\alpha$ line profiles observed with M2FS in 2022. The H$\alpha$ line is presented before and after sky correction. The best obtained fit of TLUSTY models to the H$\delta$ and He I lines are shown in dashed red.}
\figsetgrpend

\figsetend

\figsetstart
\figsetnum{A2}
\figsettitle{Hydrogen, Helium and Carbon lines identified in MIKE spectra.}

\figsetgrpstart
\figsetgrpstart
\figsetgrpnum{A2.1}
\figsetgrptitle{Target 52}
\figsetplot{full-mike-52}
\figsetgrpnote{Hydrogen, Helium and Carbon lines identified in MIKE spectra.}
\figsetgrpend

\figsetgrpstart
\figsetgrpnum{A2.2}
\figsetgrptitle{Target 56}
\figsetplot{full-mike-56}
\figsetgrpnote{Hydrogen, Helium and Carbon lines identified in MIKE spectra.}
\figsetgrpend

\figsetgrpstart
\figsetgrpnum{A2.3}
\figsetgrptitle{Target 58}
\figsetplot{full-mike-58}
\figsetgrpnote{Hydrogen, Helium and Carbon lines identified in MIKE spectra.}
\figsetgrpend

\figsetgrpstart
\figsetgrpnum{A2.4}
\figsetgrptitle{Target 68}
\figsetplot{full-mike-68}
\figsetgrpnote{Hydrogen, Helium and Carbon lines identified in MIKE spectra.}
\figsetgrpend

\figsetgrpstart
\figsetgrpnum{A2.5}
\figsetgrptitle{Target 73}
\figsetplot{full-mike-73}
\figsetgrpnote{Hydrogen, Helium and Carbon lines identified in MIKE spectra.}
\figsetgrpend

\figsetgrpstart
\figsetgrpnum{A2.6}
\figsetgrptitle{Target 75}
\figsetplot{full-mike-75}
\figsetgrpnote{Hydrogen, Helium and Carbon lines identified in MIKE spectra.}
\figsetgrpend

\figsetgrpstart
\figsetgrpnum{A2.7}
\figsetgrptitle{Target 83}
\figsetplot{full-mike-83}
\figsetgrpnote{Hydrogen, Helium and Carbon lines identified in MIKE spectra.}
\figsetgrpend

\figsetgrpstart
\figsetgrpnum{A2.8}
\figsetgrptitle{Target 84}
\figsetplot{full-mike-84}
\figsetgrpnote{Hydrogen, Helium and Carbon lines identified in MIKE spectra.}
\figsetgrpend

\figsetend

%\begin{figure}
%	\includegraphics[width=\linewidth]{figures/m2fs-1}
%	\caption{H$\delta$, He I line at 4143.76\,\r{A} and H$\alpha$ line profiles observed with M2FS in 2022. The H$\alpha$ line is presented before and after sky correction. The best obtained fit of TLUSTY models to the H$\delta$ and He I lines are shown in dashed red.}
%	\label{fig:m2fs-1}
%\end{figure}

%\begin{figure*}
%	\center
%%		\contcaption{Continued}
%	\includegraphics[width=\linewidth]{figures/m2fs-2}
%%	\caption{Fig.~\ref{fig:m2fs-1} continued}
%	\textbf{Figure A1.} Continued
%%	\label{fig:m2fs-2}
%\end{figure*}
%
%\begin{figure*}
%	\center
%	\includegraphics[width=\linewidth]{figures/m2fs-3}
%	\textbf{Figure A1.} Continued
%%	\contcaption
%%	\label{fig:m2fs-3}
%\end{figure*}
%
%\begin{figure*}
%	\center
%	\includegraphics[width=\linewidth]{figures/m2fs-4}
%	\textbf{Figure A1.} Continued
%%	\label{fig:m2fs-4}
%\end{figure*}
%
%
%\begin{figure*}
%	\center
%	\includegraphics[width=\linewidth]{figures/m2fs-5}
%%	\caption{Fig.~\ref{fig:m2fs-1} continued}
%	\textbf{Figure A1.} Continued
%%	\label{fig:m2fs-5}
%\end{figure*}
%
%\begin{figure*}
%	\includegraphics[width=\linewidth]{figures/mike-1}
%	\caption{H$\delta$, He I line at 4143.76\,\r{A} and H$\alpha$ line profiles observed with MIKE in 2021. The best obtained fit of TLUSTY models to the H$\delta$ and He I lines are shown in dashed red.}
%	\label{fig:mike-1}
%\end{figure*}
%
%\begin{figure*}
%	\center
%	\includegraphics[width=\linewidth]{figures/mike-2}
%	\textbf{Figure~\ref{fig:mike-1}.} Continued
%%	\label{fig:mike-2}
%\end{figure*}

\section{Hydrogen lines from MIKE data}

\begin{figure*}
	\centering
	\plottwo{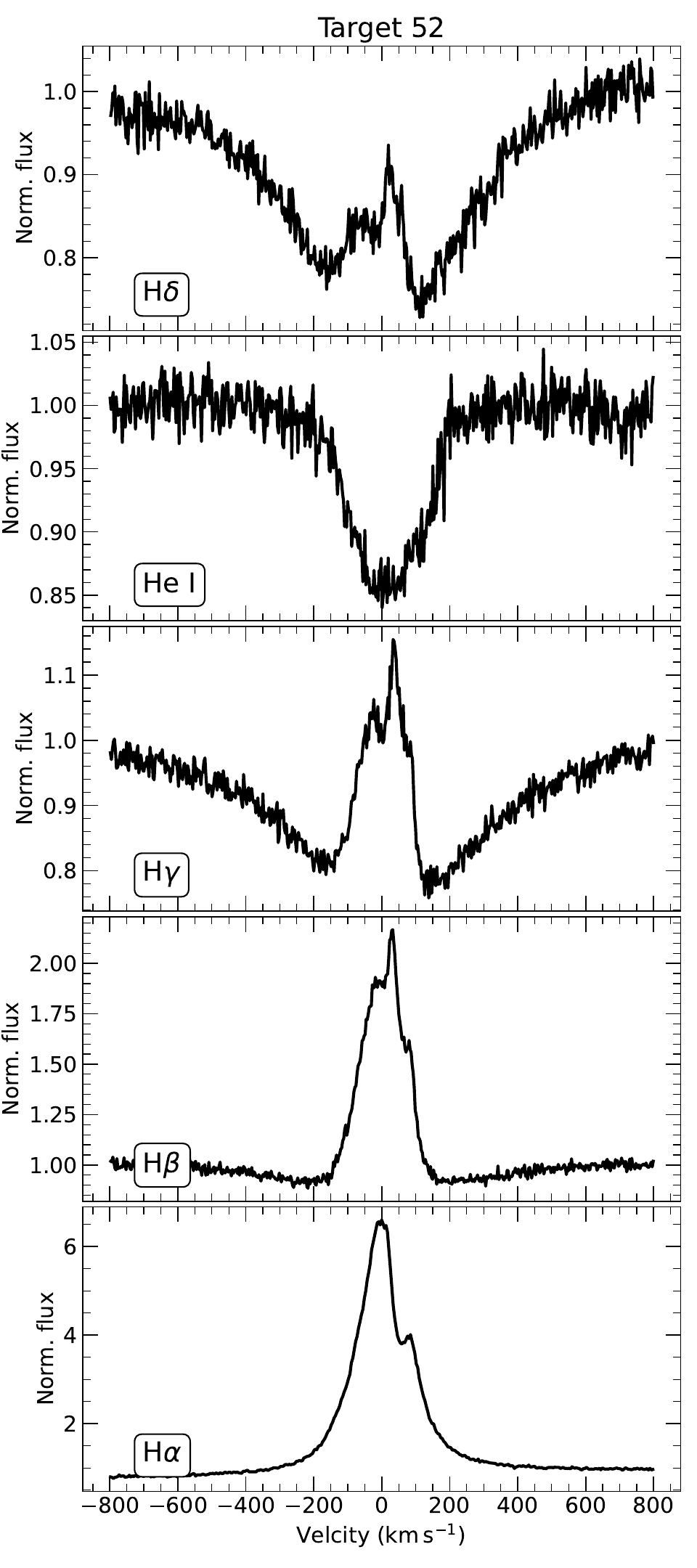}{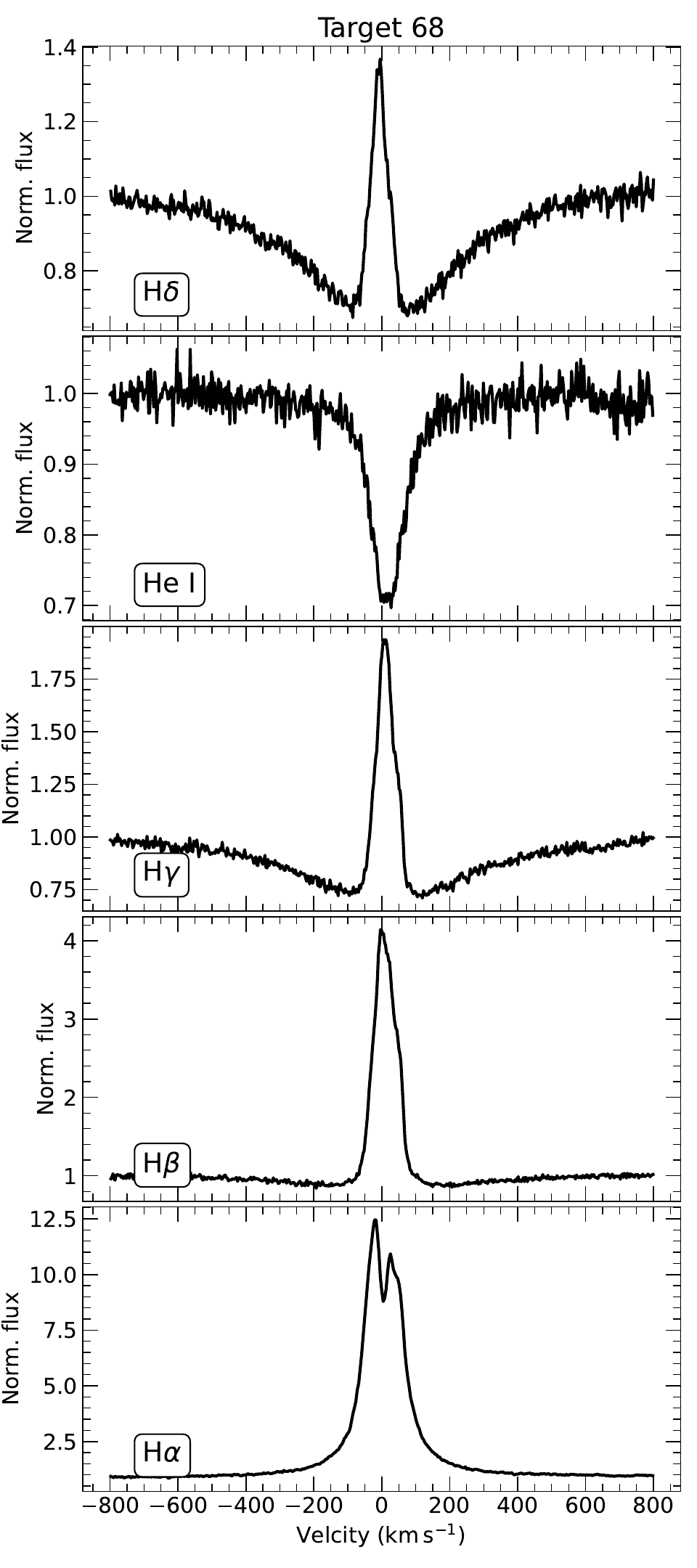}
	\caption{MIKE data for targets 52 and 68, showing the profile of the H$\alpha$, H$\beta$, H$\gamma$, H$\delta$, and the He I line at 4143.76\,$\rm \AA$}
	\label{fig:hydrogen-lines}
\end{figure*}

Figure~\ref{fig:hydrogen-lines}
%,~\ref{fig:hydrogen-lines-2} and~\ref{fig:hydrogen-lines-3} 
shows the H$\alpha$, H$\beta$, H$\gamma$, H$\delta$ and the He I line at 4143.7$\rm \AA$ for 5 targets observed with MIKE.
% Figure~\ref{fig:fwhm-hydrogen} presents the relative full width at half maximum of the hydrogen lines for three of those targets.

\figsetstart
\figsetnum{B3}
\figsettitle{H$\alpha$, H$\beta$, H$\gamma$, H$\delta$ line profiles from MIKE spectra}

\figsetgrpstart
\figsetgrpstart
\figsetgrpnum{B3.1}
\figsetgrptitle{Target 73}
\figsetplot{example-mike-73}
\figsetgrpnote{Hydrogen lines identified in MIKE spectra.}
\figsetgrpend

\figsetgrpstart
\figsetgrpstart
\figsetgrpnum{B3.2}
\figsetgrptitle{Target 83}
\figsetplot{example-mike-73}
\figsetgrpnote{Hydrogen lines identified in MIKE spectra.}
\figsetgrpend

\figsetgrpstart
\figsetgrpstart
\figsetgrpnum{B3.3}
\figsetgrptitle{Target 84}
\figsetplot{example-mike-73}
\figsetgrpnote{Hydrogen lines identified in MIKE spectra.}
\figsetgrpend

\figsetend

\bibliography{cfa.bib}{}
\bibliographystyle{aasjournal}



\end{document}